# Identifying Cocoa Pollinators: A Deep Learning Dataset

Wenxiu Xu[1,2,§], Saba Ghorbani Bazegar[2,§], Dong Sheng[1,2], Manuel Toledo-Hernández[2,3], ZhenZhong Lan[4], Thomas Cherico Wanger[2,5,6,*]

[1] College of Environmental and Resource Sciences, Zhejiang University, 310000, Hangzhou, China.

[2] Sustainable Agricultural Systems & Engineering Laboratory, School of Engineering, Westlake University, Hangzhou, China.

[3] Sustainable Development Department, Instituto Tecnológico Vale, Belém, Brazil.

[4] School of Engineering, Westlake University, Hangzhou, China.

[5] Key Laboratory of Coastal Environment and Resources of Zhejiang Province, Westlake University, Hangzhou, China.

[6] Production Technology & Cropping Systems Group, Department of Plant Production, AgroScope, Nyon, Switzerland.

[§] joined first authorship.

*Corresponding author: tomcwanger@gmail.com (TCW )

**Abstract (166 words)**

Cocoa is a multi-billion-dollar industry but research on improving yields through pollination remains limited. New embedded hardware and AI-based data analysis is advancing information on cocoa flower visitors, their identity and implications for yields. We present the first cocoa flower visitor dataset containing 5,792 images of Ceratopogonidae, Formicidae, Aphididae, Araneae, and Encyrtidae, and 1,082 background cocoa flower images. This dataset was curated from 23 million images collected over two years by embedded cameras in cocoa plantations in Hainan province, China. We exemplify the use of the dataset with different sizes of YOLOv8 models and by progressively increasing the background image ratio in the training set to identify the best-performing model. The medium-sized YOLOv8 model achieved the best results with 8% background images (F1 Score of 0.71, mAP50 of 0.70). Overall, this dataset is useful to compare the performance of deep learning model architectures on images with low contrast images and difficult detection targets. The data can support future efforts to advance sustainable cocoa production through pollination monitoring projects.

**Background & Summary**

Agricultural production has been instrumental in providing food, fiber and energy for billions of people, but environmental, health, and economic costs have rendered global farming systems a net loss system of USD 1.9 trillion annually[1]. A global food systems transformation becomes the inevitable pathway towards an agricultural future that benefits both people and planet[2,3]. Therein, biodiversity in agricultural landscapes and the associated ecosystem services such as biological pest control and pollination are critical to maintain ecosystem functioning and human wellbeing[4]. In contrast, biodiversity is also threatened by climate change, land use and pollution caused by agricultural production[5]. Effective biodiversity monitoring is, therefore, critical for global conservation efforts and to maintain ecosystem services[6] and greatly enabled through new technologies. For instance, passive acoustic monitoring and global efforts to aggregate existing information into one place can help biodiversity conservation across ecosystems[7]. Insect ecology is advancing for instance through LiDAR monitoring, DNA metabarcoding and visual methods, often coupled with machine learning approaches[8]. In general, these technologies are transforming our understanding of spatial, temporal, and taxonomic aspects of biodiversity monitoring, while it is important to overcome technical limitations and subsequent access to data bound by international standards.

Throughout the history of biodiversity monitoring, visual methods have played a pioneering role that is now advanced by AI-based methods for embedded systems. Traditional camera traps have enabled continuous documentation of animal activity during day and night. Current AI-based cameras - in particular when integrated as embedded IoT networks - can cover larger areas and collect data that can provide insights into movement patterns, species interactions, demographic trends, and behavioral patterns. Such camera systems use object detection to generate large datasets for further individual, population, and network level analyses (for laboratory studies see[9]). The general process to train object detection models is done based on established datasets or newly sourced data when specific datasets are unavailable for instance for different target organisms, lighting, or stylistic conditions. Then, Generative Adversarial Networks (GAN) can be used to generate additional data (e.g., CycleGAN can produce synthetic data[10]) or new empirical data is collected (for an example image training size effects on classification of European, see Wanger and Frohn[11]). Building new training datasets come with their own challenges. First, depending on the abundance of the target organism, obtaining enough data can be challenging. Second, the background against which the animal is captured can make the images complex and difficult to analyze[12,13]. Lastly, labelling of newly built datasets is time-consuming and expensive but can be overcome through preprocessing techniques to reduce noise in images[14].

An area where advancements in biodiversity monitoring and object detection has not been implemented is yield-determining pollination in cocoa production. Cocoa is the crop that is needed to produce chocolate, a multi-billion USD industry. It grows in the tropical regions of the world, and the crop is pollination limited[15]. When the small cocoa flowers are not pollinated, no cocoa pods emerge, and no cocoa beans can eventually be harvested leaving the cocoa industry unable to satisfy the growing global demand for chocolate. While yield benefits of manual pollination range from 200-800% in Indonesia and Brazil[16,17], little is known about natural flower visitors such as midges and flies (Dipterans), thrips (Thysanoptera), and ants and parasitoid wasps (Hymenoptera). This information, however, is critical for an effective management of cocoa plantations for pollinator conservation and to reduced environmental impacts from climate change[15,18]. Current methods for cocoa pollinator monitoring include invasive and indirect methods such as pan traps or glue that capture flower visitors but do not allow linking flower visitors to cocoa yields. Moreover, visual encounter surveys on cocoa flowers are challenging, because only 16% of all cocoa flowers are successfully pollinated and visitors spend potentially a very short time on the flowers[19]. Automated monitoring methods that can classify flower visitors are, hence, desirable because they allow linking pollination effectiveness with yields. It is, however, extremely difficult to compile the location specific datasets with flower visitors to train relevant models, because i) a low visiting frequency leads to large amounts of images without flower visitors; and ii) 24h recording requires infrared illumination for night monitoring and results in greyscale images with little contrast and sometimes only fractions of rare flower visitors.

Here, we present the first cocoa flower visitor dataset consisting of 5,792 insect images and 1,082 flower 'background' images. Of these, 5,214 insect images and 782 background images were collected in 2023, featuring five common cocoa flower visitors: Ceratopogonidae (midges), Formicidae (ants), Aphididae (aphids), Araneae (spiders), and Encyrtidae (parasitoid wasps). In 2024, 578 insect images and 300 background images were collected, featuring three common cocoa flower visitors: Ceratopogonidae, Formicidae, and Encyrtidae. The dataset was curated from 23 million images collected over two years by embedded cameras[20] deployed in cocoa plantations at the Xinglong Tropical Botanical Garden, Hainan Province, China. We then use a YOLOv8 (You Only Look Once) algorithm for object detection, predicting bounding boxes and assigning class probabilities for multiple objects in flower visitor classes[21].

**Data Records**

The full dataset can be accessed here
http://datadryad.org/stash/share/CAxX5xrwbdzlEyfW1MSR7FcNVE_H6wl29YJTuzWdyYA.

**Methods**

*Data Collection*

The data was collected in Xinglong Botanical Garden (18° 43' 57.6'' N, 110° 11' 55.8'' E), Hainan Province, China, where we monitored cocoa flowers with embedded computer vision cameras[22] for at least 24 hours. In total, we monitored 741 flowers from April to September 2023, and 417 flowers from April to July 2024. The cameras used frame differencing and blob detection to detect activity on the flowers and then automatically stored detection events on on-board SD cards. This approach resulted on average in 20,000 images per flower with and without visitor detection.

We obtained a total of 23 million images in JPG format of a 1,944 x 1,944 pixel resolution. We checked 8,040,000 images manually for flower visitors before using the trained YOLO object detection model. We employed a screening process whereby manual screening was followed up by model training and testing. We used a subset of the data to make a screening model (256 images), tested the model on new data and thereby incrementally increased the training data and performance (Fig.1A). We stopped manual screening when the model had reached 90% accuracy to save time and financial resources. Subsequently, we used the optimized model to screen for images containing insects from all the data collected in 2023. In total, we obtained 5,214 images contained flower visitors in five groups Ceratopogonidae, Formicidae, Aphididae, Araneae, and Encyrtidae from the 2023 data (Fig. 2).

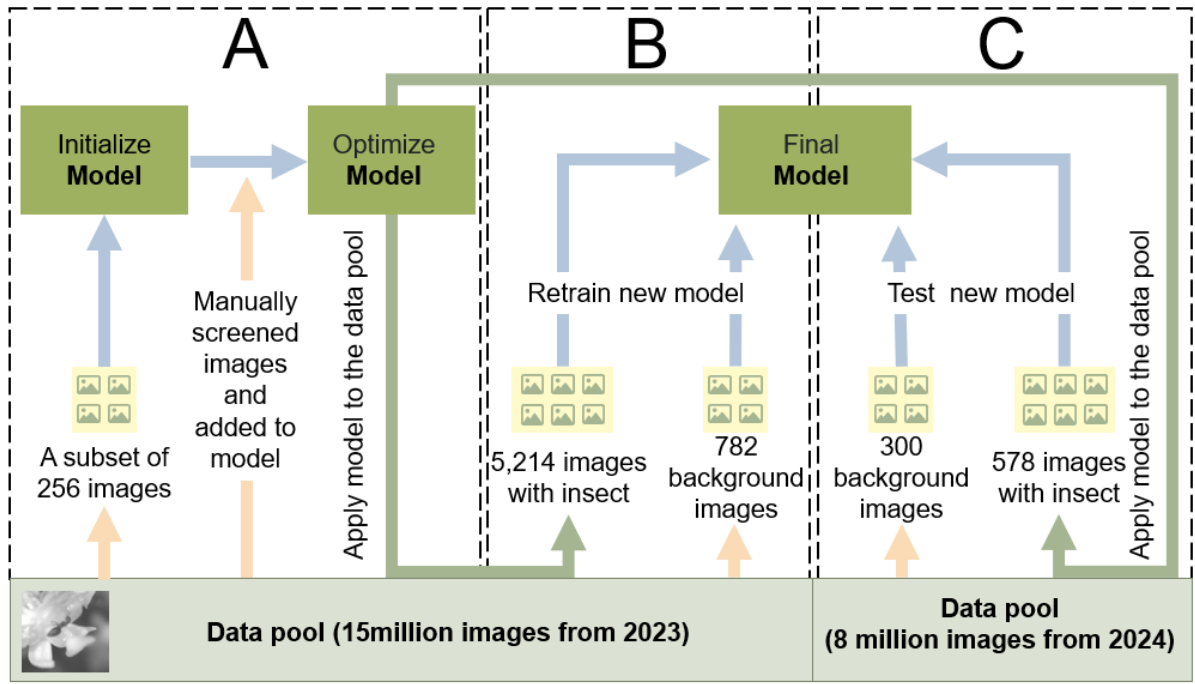

**Figure 1:** Data collection and model training process. Initial training used a subset of 256 images, followed by manual screening to optimize the model, achieving 90% accuracy (A). The optimized model was then applied to the 2023 data pool, identifying 5,214 insect images. From these, 782 unique background images were manually selected for retraining (B). In 2024, the optimized model screened 578 insect images from the data pool, with 300 unique background images manually selected to test the final model (C).

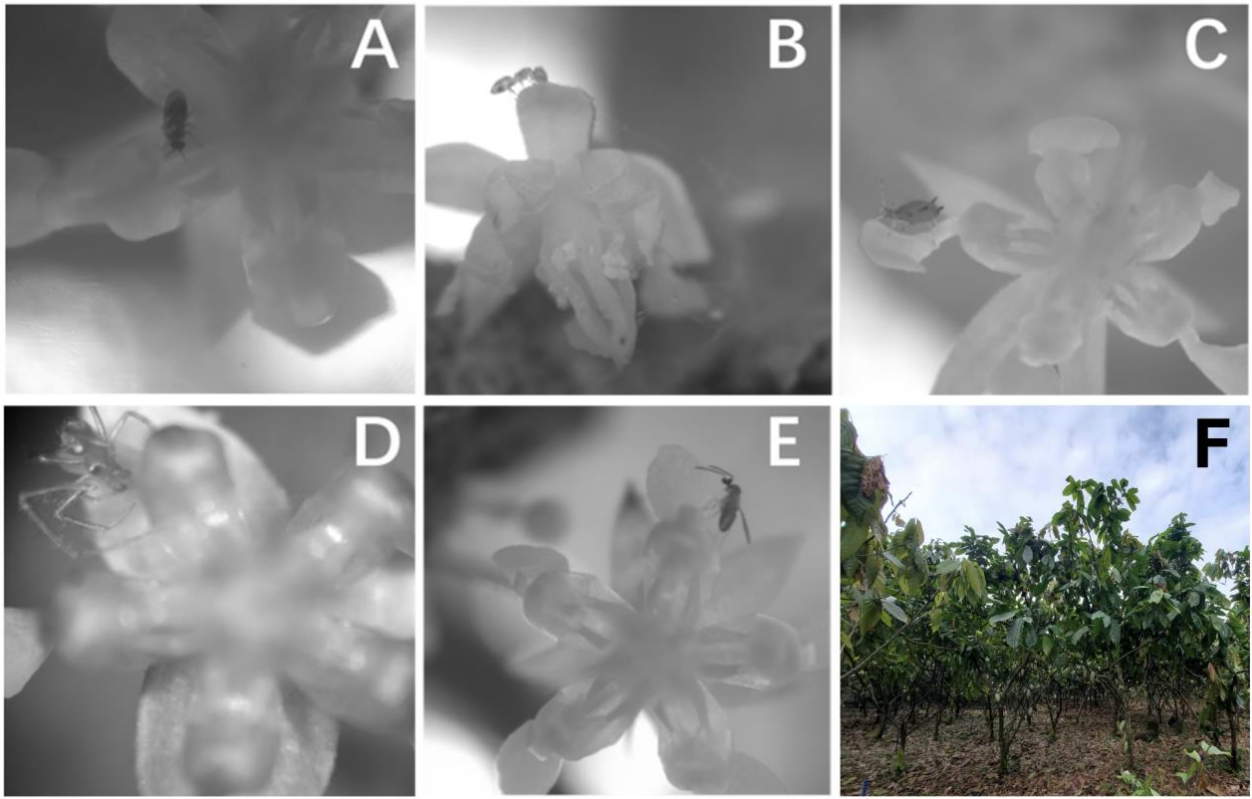

**Figure 2:** Example images for the five flower visitor groups in our dataset. Ceratopogonidae (A), Formicidae (B), Aphididae (C), Araneae (D), and Encyrtidae (E). Panel F shows the cocoa farms in Hainan, China (F; © by Manuel Toledo-Hernández).

*Object Detection Models*

Several models have galvanized as important in object detection. One such influential model is Faster R-CNN (Region-based Convolutional Neural Network) [23], which generates candidate object regions and a subsequent object detection network to classify and refine these regions. Nieuwenhuizen et al.[24] detected tomato whitefly and its predatory bugs on yellow sticky traps by a faster R-CNN model. Du et al.[25] used ResNet50 and online hard example mining to improve faster R-CNN models and detected multiple insect types in field images. The Single Shot MultiBox Detector (SSD)[26] is a popular single-stage object detection model that achieves high efficiency by simultaneously predicting object classes and bounding box coordinates at different scales using a series of convolutional layers. Lyu et al.[27] used an optimized SSD feature fusion algorithm to detect pests among grains. And Garcia et al.[28] used SSD on a microcontroller to detect and count insects such as whiteflies and aphids on eggplant leaves. Additionally, models like YOLO (You Only Look Once)[29] excel at real-time object detection, making them highly suitable for applications on embedded devices. Ratnayake et al.[30] used YOLOv4 and KNN segmentation methods to count insect visitations on a particular flower. Kumar et al.[31] introduced channel and spatial attention modules to a YOLOv5 model and detected 23 categories of insects.

We use the YOLOv8 object detection algorithm that processes input images, generates bounding boxes with corresponding class probabilities, indicating object locations and likelihoods of belonging to specific classes. YOLOv8 architecture comprises convolutional layers, spatial pyramid pooling, and Path Aggregation Network modules, enabling effective feature extraction and aggregation for accurate object detection across diverse sizes (model architecture is discussed in detail elsewhere[21];). We used the YOLOv8 model for both model creation and prediction. The YOLOv8 model consists of various weights, each with a different number of parameters. To ensure efficiency, we incorporated all these weights and compared their performance to determine which one worked most effectively.

*Dataset Annotation*

Annotating data is an important step prior to model training. It involves placing a bounding box around objects and assigning them a class for classification. For the data from 2023, we ran a preliminary YOLOv8 model on the dataset to automatically annotate objects. We then reviewed the annotations manually to delete false negatives and correcting any inaccuracies in the bounding boxes. For the data from 2024, we manually annotated it using Label-studio, and after completing the annotations, we performed a double check to ensure their accuracy. After completing the annotation of all the data, the 5,792 images contained a total of 6,027 bounding boxes, with 2,056 for Ceratopogonidae, 3,003 for Formicidae, 628 for Aphididae, 176 for Araneae, and 164 for Encyrtidae.

*Dataset Augmentation*

A general solution to limited training data that is labelled is image augmentation, whereby the images are transformed in shape and size. Augmentation techniques are dynamically applied in the training process of the model and include HSV random transformation (i.e., hue - the color tone; saturation - the intensity of color; and value – brightness of the color are modified at random). Image translation along the x & y axis introduces a position shift without structural changes and aims to train the model on different perspectives and spatial contexts. Horizontal flipping swaps pixels from one side to the other, scaling does not change the aspect ratio but object size and orientation in the frame. Lastly, mosaic augmentation uses four source images and – while preserving the aspect ratio – compiles them into a new image[32]. The original image sample size can be increased several folds, thereby enhancing generalizability of the model.

We performed horizontal and vertical flipping, image translation, and mosaic augmentation on images containing cocoa visitors. Additionally, we adjusted the brightness (V channel) in the HSV color space in our greyscale images (e.g.Fig.3). Through training data augmentation, we were able to expand our dataset 5-fold.

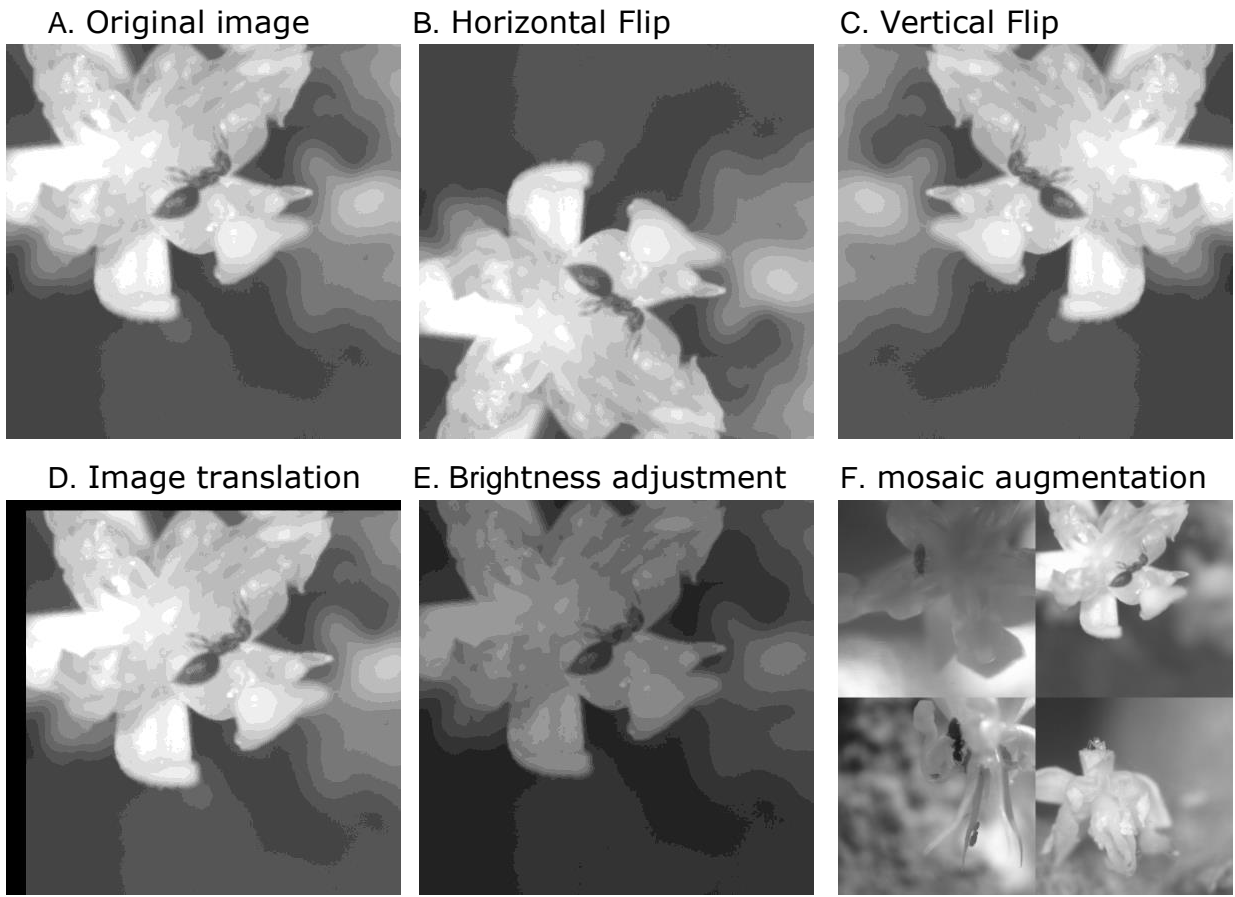

**Figure 3.** Examples of the applied image augmentation

*Experimental Setup*

The training and testing of the deep learning models were done on a workstation with the following specifications: the central processing unit (CPU) is an Intel(R) Xeon(R) W-2235 with a memory capacity of 64GB, the graphics processing unit (GPU) is an NVIDIA Quadro

RTX 4000 with 40 GB of memory; the operating system is Windows 10 Pro; the PyTorch version is 2.0.0, and the CUDA version is 11.7.

*Evaluation metrics*

We evaluated the detection performance of the model with the standard metrics precision (P), recall (R), F1 score, mAP50, mAP50-95 and false positive rate (FPR). The formulas for these evaluation indicators are as follows:

$$P = \frac{TP}{TP + FP}$$

$$R = \frac{TP}{TP + FN}$$

$$F1\ Score = 2 \times \frac{Precision \times Recall}{Precision + Recall}$$

The evaluation of insect detection is based on confidence scores. A confidence score of at least 0.5 is required to classify a flower visitor as a true positive (TP). Incorrectly identifying objects such as flower or background as a flower visitor is considered a false positive (FP). Failing to detect a flower visitor or incorrectly classifying it as a different category is considered a false negative (FN). True negatives (TN) are recorded when there is no flower visitor in the image.

Precision is the ratio of true positives to total detections made by the model, while Recall measures the proportion of true positives to the total actual objects. Additionally, we introduce the F1 score as a harmonic measure to comprehensively evaluate the model's precision and recall.

$$mAP = \frac{1}{k} \sum_{i=1}^{k} AP_i$$

Mean Average Precision (mAP) is a key evaluation metric used to assess the performance of object detection networks, as it takes both precision and recall into account. The mAP is the mean of the Average Precision (AP) values obtained at various recall levels from the Precision-Recall (PR) curve. The specific mAP50 and mAP50-95 measures the precision at an Intersection over Union (IoU) threshold of 0.50 and ranging thresholds between 0.50-0.95, respectively. The mAP50 is a measure of accuracy for 'easy' detection whereas the mAP50-95 is a more comprehensive assessment of detection performance. The IoU gives an indication of how well the predicted mask or bounding boxes match the ground truth data.

$$\text{FPR} = \frac{\text{FP}}{\text{TN} + \text{FP}}$$

The false positive rate (FPR) indicates the model's ability to distinguish cocoa visitors from background images by quantifying the rate at which background images are incorrectly identified as containing cocoa visitors.

**Technical Validation**

We used a two-step approach to build a robust and adaptable model. First, we analyzed all the data from 2023, which included 5,214 images containing identified cocoa flower visitors and 500 selected background images (Fig.1B). The model was then tested on a randomly selected subset of test data. We used 80%, 10%, and 10% of the entire dataset for training, validation, and testing, respectively. The background images were used to test the false positive rate (FPR). Due to the non-uniform distribution of data across different insect species, we used a weighted calculation method to determine the overall performance.

Second, we used the trained and validated model based on 5214 images from 2023 and tested model performance on 578 new images from 2024 (Fig.1C). The ratio of the training, validation, and testing sets is also 8:1:1. Additionally, we tested the model's false positive rate (FPR) using 300 background images collected from 2024. The testing included evaluations with different model sizes and varying proportions of background images. We repeatedly split the training and validation data sets randomly at a fixed ratio. Each model was trained three times, and the reported results are the average performance metrics from multiple test set evaluations to capture variation and model's true capacity. Our goal was to conduct adaptability tests to enhance the reliability of the model's real-world effectiveness and generalization capabilities. Due to the limited number of images from spiders and aphids in the test set, we only used three of the five classes - Ceratopogonidae, Formicidae and Encyrtidae - in our adaptability test.

The model demonstrates high detection precision (0.98), recall (0.95) and F1-score (0.96), indicating accurate and comprehensive recognition across target objects. It achieves an mAP50 of 0.97 and an mAP50-95 of 0.60, showing stable performance across various IoU thresholds. Overly optimistic detection results on test sets of different insect species are likely due to the homogeneity of the dataset. Strong internal consistency often leads to model overfitting to these patterns, high performance on the training and validation sets, but problems when deployed[33]. Furthermore, the model has a relatively high false positive rate (FPR) of 9% on background images. This issue can be resolved with more data from different locations to enhance its generalization capability of the model.

**Table 1:** Performance evaluation of the five-classes YOLOv8 model

| Class | Images | Precision | Recall | F1-score | mAP50 | mAP50-95 |
|---|---|---|---|---|---|---|
| **Overall** | 521 | 0.98 | 0.95 | 0.96 | 0.97 | 0.60 |
| Ceratopogonidae | 172 | 0.98 | 0.98 | 0.98 | 0.99 | 0.62 |
| Formicidae | 255 | 0.97 | 0.95 | 0.96 | 0.97 | 0.57 |
| Aphididae | 60 | 0.95 | 0.85 | 0.90 | 0.92 | 0.50 |
| Araneae | 22 | 1.00 | 1.00 | 1.00 | 0.99 | 0.76 |
| Encyrtidae | 13 | 1.00 | 0.92 | 0.96 | 0.96 | 0.74 |

*Background Image Addition for Model Improvement*

Background images in the training dataset can enhance object detection accuracy by enabling the model to distinguish between objects and their surroundings. We included different background images in the training dataset to avoid false negative detections. We gradually increased the percentage of background images in the training dataset from 0% to 15%, based on the 5,214 images collected in 2023.

When testing the model on images with completely unseen backgrounds, we found that training the model with an 8% background image ratio achieved the best Precision (0.78) and F1 Score (0.71) on the test set, while also yielding the lowest FPR (0.026), indicating the lowest risk of false positives. On the other hand, training with a 10% background image ratio resulted in the highest Recall (0.67), F1 Score (0.71), and mAP50 (0.74), along with a relatively low FPR (0.031). However, when the background image ratio was increased to 15%, despite achieving the lowest FPR (0.012), the overall performance of the model declined significantly (Tab. 2). Considering that real-world applications prioritize a balance in overall performance and minimal false positives, we concluded that the model trained with an 8% background image ratio is the optimal choice.

For a single class, Encyrtidae outperforms Ceratopogonidae and Formicidae in most metrics across different background image ratios (Fig. 4). For Encyrtidae, the model's performance is best when the background image ratio is 8%, achieving an F1 Score of 0.86 and an mAP50 of 0.89. For Ceratopogonidae and Formicidae, the model's F1 Score and mAP50 reach their highest values at a background image ratio of 10% (F1 Score: 0.75 and 0.65; mAP50: 0.80 and 0.66, respectively). This may indicate that the model has different sensitivity to background changes when processing different types of insects.

**Table 2:** Performance evaluation of the five-class YOLOv8 model with different background images ratio

| Background images ratio | Precision (overall) | Recall (overall) | F1score (overall) | mAP50 (overall) | mAP50-95 (overall) | FPR |
|---|---|---|---|---|---|---|
| 0% | 0.74 | 0.65 | 0.69 | 0.69 | 0.30 | 0.041 |
| 5% | 0.72 | 0.636 | 0.67 | 0.67 | 0.31 | 0.033 |
| 8% | 0.78 | 0.65 | 0.71 | 0.70 | 0.31 | 0.026 |
| 10% | 0.77 | 0.67 | 0.71 | 0.74 | 0.33 | 0.031 |
| 12% | 0.72 | 0.59 | 0.64 | 0.65 | 0.307 | 0.013 |
| 15% | 0.74 | 0.63 | 0.68 | 0.67 | 0.30 | 0.012 |

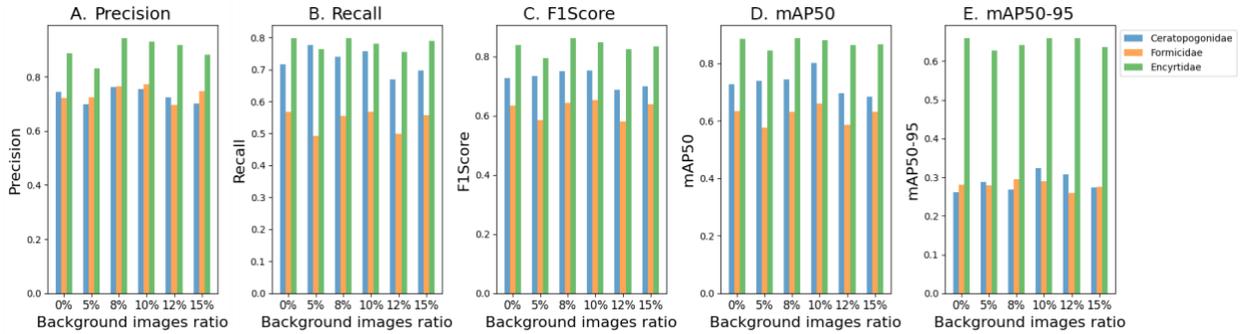

**Figure 4:** Evaluation metrics for the adaptability test of the optimal models trained with background image proportions increasing from 0%, 5%, 8%, 10%, 12%, to 15%.

*Analysis Based on Model Size*

Model size is mostly a trade-off between detection accuracy and computational demand, which must be evaluated for each application. In some cases, using a very large model may not necessarily result in higher accuracy. This phenomenon, known as underfitting, occurs when the model is too complex for the available data. To determine the optimal accuracy and model size, we conducted a series of tests on our dataset using different sizes of the YOLO model (for a detailed description of the YOLO model refer to[34]). Specifically, we evaluated the performance of YOLOv8 models with increasing complexity and size from YOLOv8n, YOLOv8s, YOLOv8m, and YOLOv8l. Our experiments aimed to identify the model that provides the best balance between accuracy and computational efficiency.

Performance and training effectiveness improved in all models as the number of training epochs increased (Fig. 5). Among them, the smaller models, YOLOv8n and YOLOv8s, converged faster, while the larger models, YOLOv8m and YOLOv8l, had a slower convergence rate but ultimately achieved lower loss values. From the perspective of training performance, the performance of YOLOv8n, YOLOv8s, and YOLOv8m are similar (Fig. 6). This suggests that a larger model size does not necessarily lead to better training results with our dataset.

When evaluating pre-trained YOLOv8 models on test images that were unseen during training, YOLOv8m demonstrated superior overall performance. It achieved the highest Precision (0.78), Recall (0.65), F1 Score (0.71), and mAP50 (0.70) among all tested models (Tab. 3).

As the model size increases, the false positive rate (FPR) for background images shows a gradual upward trend. This is attributed to the increased model complexity and the higher number of network parameters in YOLOv8m and YOLOv8l, which enhances the models' ability to extract detailed features—especially for small objects and complex scenes. However, this improvement comes at the cost of a higher propensity to generate false alarms from intricate background features, leading to an elevated FPR. Despite this trade-off, YOLOv8m strikes a better balance between accuracy and false positive rate, making it a strong candidate for applications requiring both precision and robustness in complex environments.

In single-category detection, YOLOv8m performs better in the detection of Formicidae and Encyrtidae. Ceratopogonidae exhibited the best performance on YOLOv8s (Fig. 7). This indicates that the YOLOv8m model offers the most optimal performance in detecting visitors to cocoa flowers based on our dataset. Despite the smaller number of Encyrtidae images, the insects in these images were clearly visible and displayed distinct features. This clarity allowed the model to perform well even on images it had not seen during training. In contrast, the images of Ceratopogonidae were more numerous but often suffered from poor focus, leading to incomplete or indistinct insect shapes, sometimes appearing as mere black dots. This made it difficult for the model to distinguish the insects from the background. For Formicidae, the available images contained different species of varying sizes, which made it harder for the model to generalize effectively.

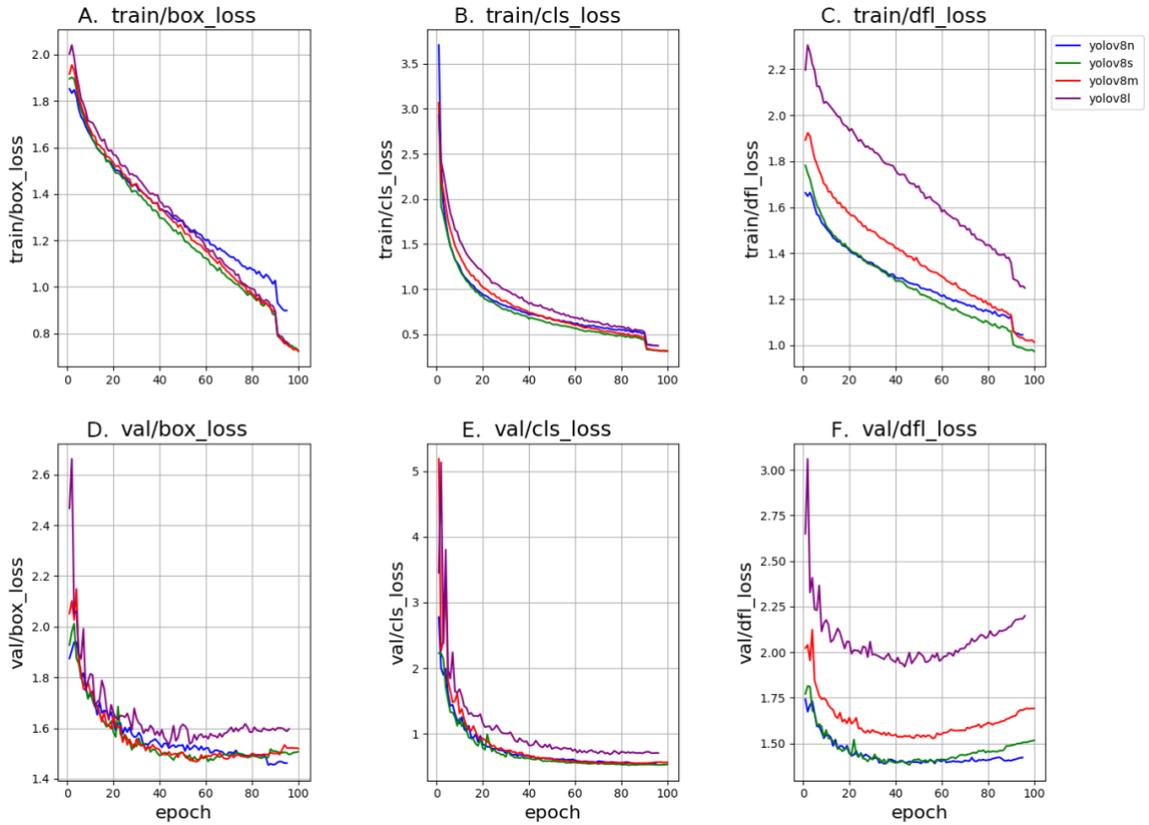

**Figure 5.** Comparison of Loss Curves during training for models of different sizes. A. train/box loss= Localization error for predicted vs. ground truth boxes during training; B. train/cls_loss= Classification error during training; C. train/dfl_loss= Focal loss optimizing bounding box regression during training; D. val/box loss= Localization error during validation; E. val/cls_loss= Classification error during validation; F. val/dfl_loss= Focal loss for bounding box regression during validation

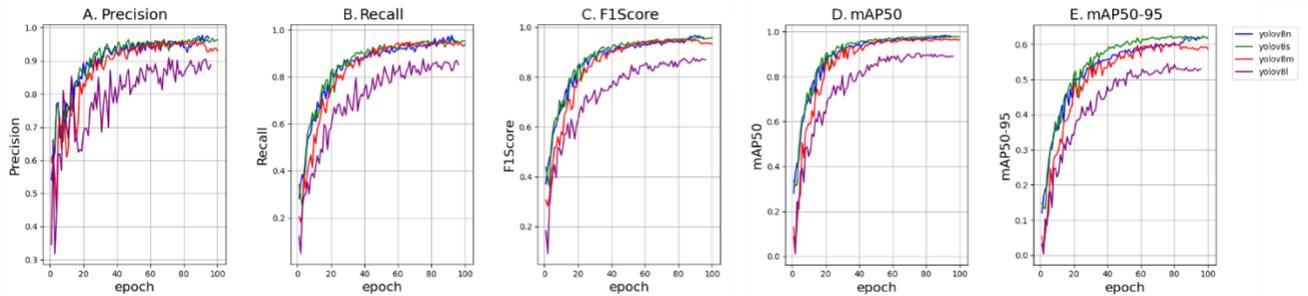

**Figure 6.** Comparison of performance metrics (A. Precision, B. Recall, C. F1 Score, and D&E Mean Average Precision (mAP)) for YOLOv8 models of different sizes during training.

**Table 3:** Performance evaluation of YOLOv8 models of different size

| Model | Params(M) | Precision (overall) | Recall (overall) | F1score (overall) | mAP50 (overall) | mAP50-95 (overall) | FPR |
|---|---|---|---|---|---|---|---|
| YOLOv8n | 3.2 | 0.72 | 0.55 | 0.61 | 0.63 | 0.27 | 0.016 |
| YOLOv8s | 11.2 | 0.77 | 0.60 | 0.67 | 0.68 | 0.29 | 0.022 |
| YOLOv8m | 25.9 | 0.78 | 0.65 | 0.71 | 0.70 | 0.31 | 0.026 |
| YOLOv8l | 43.7 | 0.75 | 0.63 | 0.68 | 0.65 | 0.28 | 0.027 |

Reported are means from three model runs for each YOLO size class.

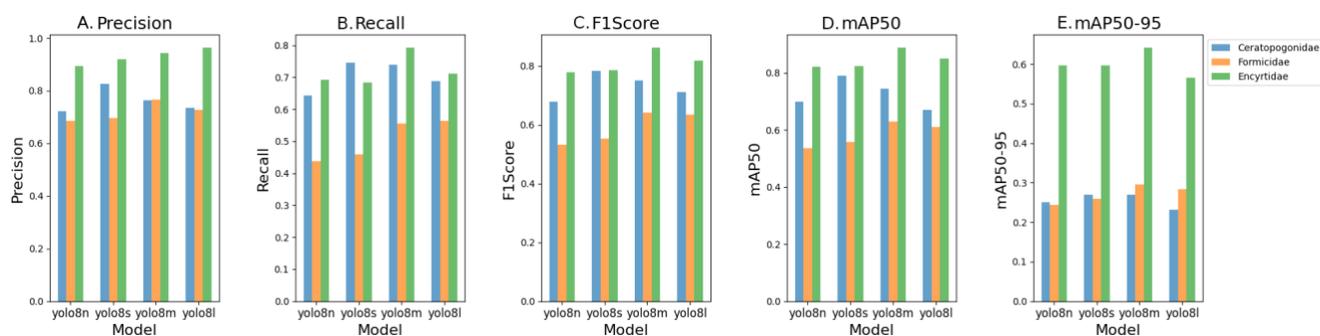

**Figure 7.** Performance comparison of YOLOv8 models of different sizes on the test dataset.

This work shows that parameter choice, and the percentage of background image inclusion was critical to enhance model performance to detect economically viable cocoa flower visitors. A medium-sized YOLOv8 model with 25.9 million parameters, trained on a dataset with 8% background images, achieved the best performance in recognizing three categories. The model attained a Precision of 0.78, Recall of 0.65, F1 Score of 0.71, and mAP50 of 0.70, with a false positive rate of 2.6%. These results suggest that to identify cocoa flower visitors under challenging field conditions, it is critical to increase the amount of data for each class to be detected and, therefore the total number of field deployments. To further enhance detection accuracy across different environments, future efforts could focus on optimizing detection algorithms or increasing the diversity of insect images. This work provides a foundational basis for advancing AI-driven solutions in the cocoa farming industry.

**Code Availability**

No custom code was used to generate or process the data described in the manuscript.

**Acknowledgements**
We would like to thank Professor Li Fupeng from the Spice Beverage Research Institute of the Chinese Academy of Tropical Agricultural Sciences for his support in data collection. We also thank students in Xinglong botanical garden for their help with data collection in 2023 and 2024. This work was funded through a Westlake University Startup Fund to TCW.

**Author Contributions**
Conceptualization: WX, SGB, ZL, TCW; Data collection: WX, SGB, MTH; Methodology and analyzes: WX, SGB, ZL, TCW; Visualization: WX, DS, MTH; Writing – original draft: WX, SGB, DS, TCW; Writing – review & editing: all authors; Funding acquisition, Project administration & Supervision: TCW.

**Competing Interests**
The authors do not have competing interests.